\documentclass[preprint2]{aastex}

\newcommand {\belist}{\begin{list}{---}{\setlength{\rightmargin}%
{\leftmargin}}}

\newcounter{arabnum}
\newcommand{\befigcap}{\begin{list}{ {\bf Figure \arabic{arabnum} } %
{ \usecounter{arabnum}} } }
\newcommand{\enfigcap}{\end{list}}

\newcommand{\bequo}{\begin{quotation}}
\newcommand{\enquo}{\end{quotation}}
\newcommand{\bverse}{\begin{verse}}
\newcommand{\everse}{\end{verse}}
\newcommand{\beit}{\begin{itemize}}
\newcommand{\enit}{\end{itemize}}
\newcommand{\been}{\begin{enumerate}}
\newcommand{\enen}{\end{enumerate}}
\newcommand{\ecen}{\end{center}}
\newcommand{\bcen}{\begin{center}}

\newcommand{\begeq}{\begin{equation}}
\newcommand{\eneq}{\end{equation}}
\newcommand{\befig}{\begin{figure}}
\newcommand{\enfig}{\end{figure}}

\newcommand{\ferrh}{\mbox{$\rm{[Fe/H]}$}\ }

\newcommand{\kmsec}{\mbox{${\rm \: km\:s^{-1}}$}\ }

\newcommand{\msol}{\mbox{$\: M_\odot$}\ }

\newcommand{\yrn}{\mbox{${\rm \:yr^{-1}}$}\ }

\newcommand{\uvby}{\mbox{\it uvby }\ }

\begin{document}
 
%
%
%
%
%
%
%
%
%
 
\title{ Age difference between the populations of binary and single
F~stars revealed from {\it HIPPARCOS}  data}
 
\author{A.A. Suchkov}

\affil{Space Telescope Science Institute\altaffilmark{1}\altaffiltext{1}{
Operated by AURA Inc., under contract with NASA, Baltimore, MD 21218, USA} \\
Received {\it February 7, 2000 }; accepted {\it April 21, 2000 }
}
\begin{abstract}

We have compared the kinematics and metallicity of the main sequence 
binary and single \uvby\ F~stars from the {\it HIPPARCOS}  catalog 
to see if the populations of these stars 
originate from the same statistical ensemble. 
The velocity dispersions of the known unresolved binary F~stars 
have been found to be dramatically smaller  than those of the single F~stars. 
This suggests that the population of these binaries is, in fact,
younger than that of the single stars, which is further supported by 
the difference in metal abundance: the binaries turn out to be, 
on average, more metal rich than the single stars. So, we conclude that 
the population of these binaries is indeed 
{\it younger} than that  of the  single F~stars. 
Comparison of the single F~stars with the C~binaries (binary candidates 
identified in Suchkov \&\ McMaster 1999) has shown, on the other hand,
that the  latter stars are, on average, {\it older} than 
the single F~stars. We suggest that  the age difference between 
the single F~stars, known unresolved binaries, and C~binaries is 
associated with the fact that  stellar evolution in a binary 
systems depends on the binary components mass ratio and separation,
with these parameters being statistically very different for
the known binaries and C~binaries (e.g., mostly substellar secondaries
in C binaries versus stellar secondaries in known binaries). 

In general we conclude that the populations of known binaries, C binaries,
and single F~stars do not belong to the same statistical ensemble.
The implications of the discovered age difference between these populations 
along with the corresponding differences in kinematics and metallicity
should be important not only for  understanding the evolution 
of stars but also for the history of star formation and the evolution of the 
local galactic disk.

\end{abstract}

\keywords{binaries: general ---  Galaxy: kinematics and dynamics --- 
 Galaxy: stellar content --- solar neighborhood --- stars: evolution --- stars: kinematics
}

\section{\bf Introduction}

Binaries are believed to constitute a significant fraction of 
the local stellar population (e.g., Duquennoy A. \&\ Mayor 1991, 
Suchkov \&\ McMaster 1999). Because of that they contribute substantially 
to the statistics of age, metallicity, and kinematics  of the 
main sequence stars in the solar neighborhood.
Since the evolution of the local galactic disk is deduced
mostly from these statistics (e.g., Twarog 1980, Carlberg et al. 1985,
Marsakov et al. 1990, Meusinger et al. 1991, Edvardsson et al. 1993,
Caloi et al. 1999), the presence of binaries may affect 
inferences regarding the history of star formation and metal enrichment
in the disk if the population of these stars is statistically different 
from that of the single stars. 
In this paper we compare the kinematics and metallicity of binary and 
single F~stars to see if they belong to the same statistical ensemble.

We use a sample of the {\it HIPPARCOS} F~stars ($0.22 \geq (b-y) \geq 0.39$)
that have \uvby photometry in Hauck \&\ Mermilliod (1998),
and treat separately the stars marked in {\it HIPPARCOS}  as
single stars, unresolved  binaries, and resolved binaries.
The latter two groups of stars will be referred to as
{\it known} binaries to differentiate them from the binary candidates
(C binaries) discussed below. 
All the stars have tangential velocity obtained
from  the {\it HIPPARCOS}  parallax and proper motion.
About a third of them have also radial velocities (taken mostly from
Ochsenbein F. 1980, Barbier-Brossat et al. 1994, and Duflot et al. 1995), 
which have been used to compute total spatial velocity. 
Metallicity has been  derived from the \uvby data using the metallicity
calibration from Carlberg et al. (1985). Effective temperature, $T_e$,
has been computed from the $(b-y)$ color, with 
the algorithm based on Moon (1985),  Moon \&\ Dworetsky (1985).
The sample is constrained to the stars within 200~pc, 
metallicity range $-0.6 < \ferrh < 0.5$, magnitude accuracy better 
then 0.003 (in $V$), parallax accuracy better than 3 mas, 
and proper motion accuracy better than 3 mas~\yrn in both coordinates.

\begin{figure}[hp]
\plotone{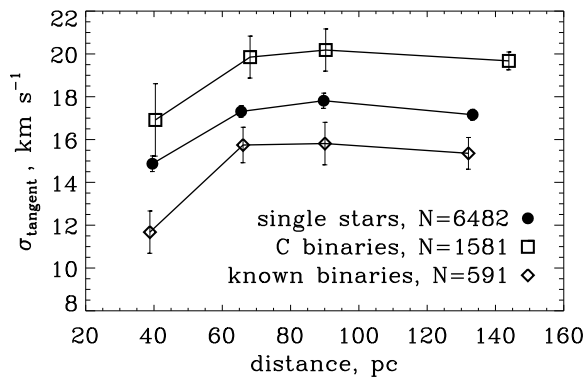} \caption{Kinematics of F~stars as a function of distance. 
The difference in tangential velocity dispersion between the three groups 
of stars suggests that the known binaries  are  statistically
younger while the C binaries are older than the single F~stars. 
\label{fig-knm} }
\end{figure}

Many of the nearby  single stars are 
in fact yet unidentified binaries (e.g., Duquennoy \&\ Mayor 1991).
Numerous binary candidates, called C~binaries, were identified  
in Suchkov \&\ McMaster (1999). The criterion used to isolate C~binaries 
involves the difference between the star's absolute 
magnitude derived from the {\it HIPPARCOS}  parallax and the Johnson 
$V$~magnitude, $M_V$, and the absolute magnitude computed from the dereddened
\uvby color index $c_0$, $M_{c_0}$ (the computation of $M_{c_0}$ utilizes
the algorithm based on Moon 1985, Moon \&\ Dworetsky 1985). 
The discrepancy between the two magnitudes, 
$\Delta M_{c_0} = M_{c_0} - M_V > 0$ ($M_V$ too bright for the star's 
$\Delta c_0$), appears to be a strong signature of a binary system.
So, to reduce the contamination in the sample of the single stars with
binaries, we have excluded the stars with $\Delta M_{c_0} > 0.45$
(C~binaries at the 3-$\sigma$ level) from that sample. 
The same rejection criterion 
has been applied to the sample of the binary stars to avoid any 
differential bias. The anomalously bright ``single'' stars, 
$\Delta M_{c_0} > 0.45$, have been isolated into a sample of C~binaries.
A summary of kinematics and metallicity of all groups of stars
considered here is given in Table~1.

\begin{table}

\caption{Velocity dispersion and mean metallicity of single and 
binary F~stars within 200 pc. 
}
\begin{tabular}{lcccr}\\ \tableline \tableline
 
 Sample   & $\Delta M_{c_0}$ & $\sigma$ & [Fe/H] &  N   \\
 {}      &  (mag)    &        (\kmsec)  & ($\times 10^{-2}$) & {} \\ \tableline

& \multicolumn{4}{c}{tangential velocity stars, $\sigma=\sigma_{tangent}$ } \\
         
 single\tablenotemark{a} \dotfill 
         & $<0.45$ & 17.1 $\!\pm\!$ 0.2 & $-1.09 \!\pm\! 0.2$ &  6482 \\
 binary\tablenotemark{b} \dotfill 
         & $<0.45$ & 15.2 $\!\pm\!$ 0.4 & $-0.75 \!\pm\! 0.7$ &   591 \\
          
``single''\tablenotemark{c} \dotfill 
         & $>0.45$ & 19.7 $\!\pm\!$ 0.3 & $-1.06 \!\pm\! 0.5$ &  1581 \\
 binary \dotfill  
         & $>0.45$ & 17.0 $\!\pm\!$ 0.5 & $-1.02 \!\pm\! 0.7$ &   641 \\

& \multicolumn{4}{c}{spatial velocity stars, $\sigma=\sigma_{total}$ } \\

 single \dotfill  
         & $<0.45$ & 22.0 $\!\pm\!$ 0.4 & $-1.09 \!\pm\! 0.5$ &  1414 \\
 binary \dotfill  
         & $<0.45$ & 15.9 $\!\pm\!$ 0.9 & $-0.65 \!\pm\! 0.3$ &   147 \\

``single''\tablenotemark{c} \dotfill
         & $>0.45$ & 23.4 $\!\pm\!$ 0.9 & $-1.02 \!\pm\! 0.1$ &   332 \\
 binary \dotfill  
         & $>0.45$ & 17.0 $\!\pm\!$ 0.8 & $-0.97 \!\pm\! 0.3$ &   203 \\
\end{tabular}
\tablenotetext{a, b}{used to generate plots for single and known binary 
stars, respectively}
\tablenotetext{c}{called C~binaries in the text}
 
\end{table}

\section{\bf Evidence from kinematics and metallicity
for age difference between binary and single F~stars}
Figure~\ref{fig-knm} compares tangential velocity dispersion of the 
single F~stars, 
known unresolved binaries, and C binaries, revealing a dramatic 
difference between the three populations. First, the known binary F~stars have
significantly smaller velocity dispersions at all distances from the Sun.
The latter fact rules out a possible selection bias against distant binaries
as a cause of the kinematic differences between the three populations. 
As seen in Table~1, the difference in spatial velocity dispersion
is even more dramatic.

\begin{figure}[hp]
\plotone{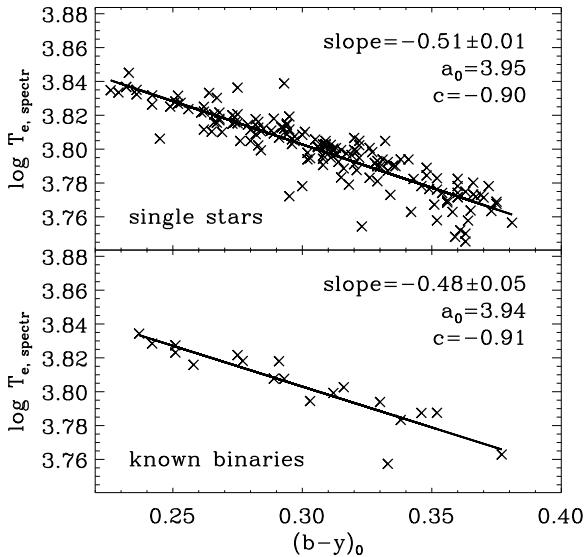} \caption{ Relationship between the temperature measured 
from spectroscopic analysis, $T_{e, spectr}$, and the color index
$(b-y)_0$.  The legend gives
the  correlation coefficient, $c$, along with the intercept, $a_0$, and
the slope of the regression line. The numbers show that the relationship 
for the binaries (lower panel) is essentially identical to that for the 
single stars (upper panel), which means that the $(b-y)_0$ color selection 
criterion provides the same age sampling for binary and single stars.
\label{fig-tef} }
\end{figure}

We have checked if the sample color selection criterion, 
$0.22 \leq (b-y)_0 \leq 0.39$, introduces a bias  favoring
hotter binaries and/or cooler single stars, which could make the sample 
binary stars younger just because of different age sampling at different 
effective temperatures.  Such a bias can arise if, for whatever reasons,
binaries are hotter than single stars at the same $(b-y)_0$ color.
If this is the case, the relationship between the effective temperature,
$T_{e, spectr}$ (derived from spectroscopic analysis),
and  $(b-y)_0$ must be  different for the  binaries and the single stars. 
This is tested in Figure~\ref{fig-tef}, where $T_{e, spectr}$ 
is from Cayrel de Strobel et al. (1997).  Comparison 
of the upper and lower panels shows that within the uncertainties
the two relationships are identical, which rules out any age bias 
associated with the color selection criterion.
Among  other known selection effects none was found  to introduce 
a bias favoring low-velocity binary stars.  So, the kinematic 
discrepancy between the known binaries and the single stars in 
Figure~\ref{fig-knm} must be intrinsic.   

\begin{figure}[hp]
\plotone{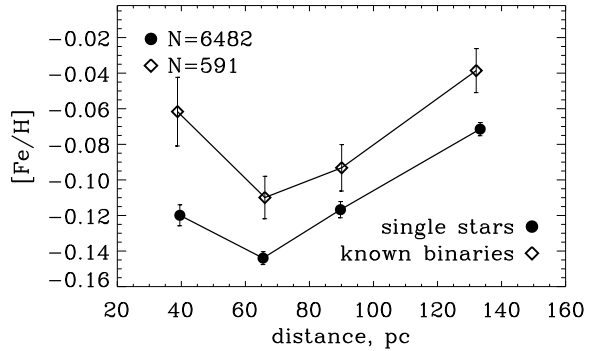} \caption{ Metallicity  of known binary
and single  F~stars as a function of distance.
The higher metallicity of the binary stars indicates that they are
younger than the single stars.
\label{fig-feh} }
\end{figure}

For the stellar populations in the  solar neighborhood,  smaller velocity 
dispersion implies younger age. Therefore, the above result means that 
the known binary F~stars are, on average, younger than single F~stars.  
This conclusion is further supported by the respective discrepancy in the
mean metallicity of the two populations. Figure~\ref{fig-feh} shows that
the binary stars are, on average, more metal rich,
implying that they are statistically younger than single stars.  
As with the kinematics, the difference in metallicity is present at all 
distances, and  none of the known selection effects was found  
to introduce a bias favoring metal rich binary stars.
This argues that the metallicity difference and the corresponding
age discrepancy between the known binaries  and the single stars are real.

\begin{figure}[hp]
\plotone{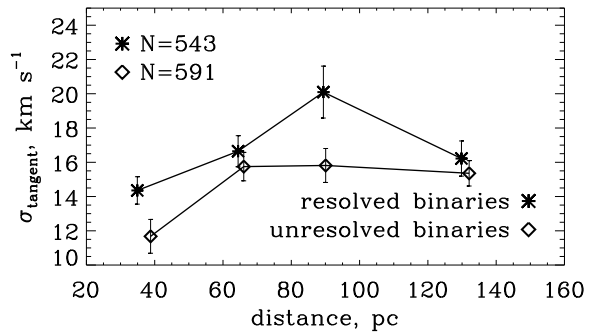} \caption{ Kinematic discrepancy between the spatially 
resolved  and unresolved  binary F~stars.
Larger velocity dispersions (older age) of resolved binaries 
apparently are due to larger component separation,
which suggests that the lifetime of the main sequence binary F~stars  
depends on component separation. 
\label{fig-i1i2mh} }
\end{figure}

If the population of known binaries is indeed younger than 
that of single F~stars, the questions is: why?
An obvious place to look for an answer is the interaction
of the binary components. 
Depending on separation, the components of close binaries can 
strongly interact with each other, including mass loss and/or mass transfer, 
which drastically changes their evolution. The interaction effects 
are believed to make the CM diagrams of star clusters look much different 
from what is expected on the basis of the standard stellar evolution
theory as applied to a coeval population of single stars (see, e.g., 
Pols \&\ Marinus 1994, and references therein). So, it would be natural
to associate the younger, on average, age of the binary F~stars with the
interaction of binary components in tight pairs, which somehow reduces 
the main sequence lifetime of the primary component. 
This hypothesis is supported by Figure~\ref{fig-i1i2mh},
which  provides evidence that component separation 
is relevant to  how long the primary star may live on the MS.   
Figure~\ref{fig-i1i2mh} compares the velocity dispersions of the spatially  
resolved binary F~stars (in fact, these are the binary components having  
individual entries in the {\it HIPPARCOS} catalog) with those of the
unresolved double stars. Statistically, the spatially resolved binary 
pairs have obviously wider separations than the unresolved binaries. 
Therefore, if there is any impact  of component interaction on stellar 
evolution, it must be less significant among the resolved binaries,
so the populations of resolved and unresolved binaries must exhibit 
statistical differences arising from different stellar evolution
in wide and tight binary pairs.
Figure~\ref{fig-i1i2mh} shows that  this is indeed the case.
The kinematics of the unresolved binaries turns out to be younger
(smaller velocity dispersion) than that of the resolved binaries at any 
distance from the Sun, indicating that binary pairs with smaller component
have shorter lifetimes.  

\begin{figure}[hp]
\plotone{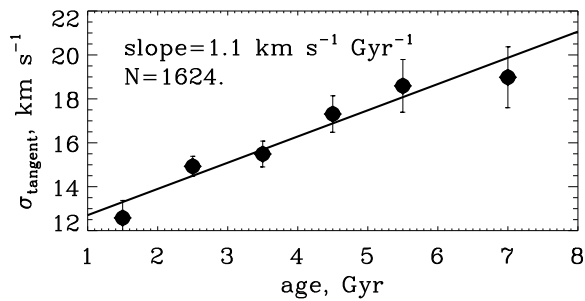} \caption{Age--velocity relation for 
F~stars within the range of $\Delta M_{c_0}$  occupied
by the best known nearby single stars, $\Delta M_{c_0}=\pm 0.15$. 
The slope of the linear regression  implies  that a 
difference of $\sim 1 $~\kmsec in tangential velocity dispersion corresponds
to the age difference of $\sim 1$~Gyr.
\label{fig-avr} }
\end{figure}

A crude estimate of the age difference between the binary and single
F~stars can be obtained using age--velocity relation (AVR). 
To illustrate this, we have derived 
the AVR for the subsample of F~stars 
constrained to the range of $\Delta M_{c_0}$ shown by the best
known single stars within 25~pc, $\Delta M_{c_0} \pm 0.15$,
and  metallicity range $-0.3 \leq \ferrh \leq 0.01$.
The isochrones for isochrone fitting are from Demarque et al. 
1996\footnote{Available at http://shemesh.gsfc.nasa.gov/astronomy.html.}, 
for the composition matching the metallicity 
range of the selected stars: $Z=1\times 10^{-2}$,  $ Y=0.27$.
This AVR is shown in Figure~\ref{fig-avr}. The slope of the linear 
regression in Figure~\ref{fig-avr} indicates that a 1~\kmsec difference 
in tangential velocity dispersion, $\sigma_{tangent}$, corresponds to a 
$\sim 1$~Gyr difference in average age. The unresolved 
binaries have $\sigma_{tangent}$ about $ 2$ \kmsec\ smaller than 
$\sigma_{tangent}$
of the single stars (Table~1), which means that they  are statistically
{\it younger} by at least $\sim 2$~Gyr.

In contrast to the known binaries, the C~binaries  
turn out to be kinematically {\it older} than the single F~stars. 
One can infer this from Figure~\ref{fig-knm} and Table~1, 
where higher velocity dispersions of the C~binaries indicate older age. 
From the slope of the AVR in Figure~\ref{fig-avr} and  the average difference 
in tangential velocity dispersions of $\sim 2$~\kmsec (Table~1), we 
conclude, that the age difference between the two population is $\sim 2$~Gyr.

There is evidence that the older age of C~binaries is associated
with retarded stellar evolution of a primary F~star in a binary
system, with $\Delta M_{c_0}$  being a measure of retardation
(Suchkov 1999); the retardation probably results from component interaction 
in tight binary pairs, which is suggested by a strong correlation 
between $\Delta M_{c_0}$ and binary components separation found 
for known binaries.
However, if C~binaries are double star systems similar 
to known binaries, the question is, why they evolve differently.
We believe that the solution to this puzzle will eventually be found in 
the details of interaction in tight binary pairs. 
The outcome of the binary components interaction 
in terms of its impact on stellar evolution should depend on the combination
of the binary system parameters, such as the component mass ratio, 
separation, orbit eccentricity, etc. It seems plausible that depending on 
that combination, the interaction of binary components  
can alter the standard evolution of the main sequence primary 
either by affecting the hydrogen-burning core
and prolonging the star's main sequence lifetime,  
as implied in some scenarios of blue straggler formation 
(cf. Wheeler 1979; see also Livio 1991), or by various mass loss/transfer 
events that interrupt the normal main sequence evolution 
(e.g.,  Pols \&\ Marinus 1994) or both.
The old age of C~binaries suggests that the former mechanism 
statistically dominates the evolution of these stars. One can speculate that 
this is because the secondary in many of these systems 
is probably too small (substellar)
to evoke appreciable mass loss/transfer\footnote{This would also explain 
why these stars have been so successful in escaping from being detected 
as binary stars.}
but is very close to the primary to tidally induce instabilities 
leading to enhanced internal mixing, which provides a fresh supply of
hydrogen to the core\footnote{A corollary would be that 
C~binaries may turn out to be the best targets to look for 
existence of close substellar/planet companions.}. 

In the case of known  binaries, both components  are stellar,
so both of the above mechanisms can operate in these systems. 
This means that although stellar evolution in a tight pair can be
slower, the primary can prematurely  cease to exist as a main
sequence F~star because of a strong mass loss occurring at some point 
in the evolution of the binary.

In fact, the latter scenario predicts that the known binaries with 
far evolved primaries should be, 
on average, older than expected from the standard stellar evolution 
models.  This prediction can be verified by comparing 
velocity dispersions of binaries with different $\Delta M_{c_0}$.
As mentioned above, $\Delta M_{c_0}$ appears to provide a measure of stellar 
evolution retardation. The larger $\Delta M_{c_0}$  the larger
amount of extra time spent by the star on the  main sequence,  so
the stars with large $\Delta M_{c_0}$  must be statistically
older than those with small $\Delta M_{c_0}$. 
This seems to be indeed the case. 
Based on the age--velocity relation in
Figure~\ref{fig-avr}, the difference in tangential velocity dispersion
of $1.9$~\kmsec between the two groups of known binaries (see Table~1)
implies that those with a strong signature of retardation,
$\Delta M_{c_0} > 0.45$,  are older (statistically) by 
$\sim 2$~Gyr\footnote{Retarded stellar evolution 
in double stars finds support in recent studies of eclipsing binaries 
with accurate sizes of their components.
It turns out that the isochrone age of the evolved primary star in 
an eclipsing binary is typically much younger than that of the
less evolved  secondary (Popper 1997; see also Clausen et al. 1999, and
Gimenez et al. 1999). Unless the current stellar evolution models are 
totally wrong below $\sim 1 \msol$, this discrepancy is an indicator 
that such a primary evolves at a  much slower pace than expected, 
so the  standard isochrones underestimate its age.}.

In summary, we conclude that the populations of binary and single F~stars 
do not belong to the same statistical ensemble.  The main difference is
in the mean age of the populations, which has  important implications 
for the modeling of both stellar evolution and galactic evolution.

\acknowledgements{
It is a pleasure to thank 
L. Yungelson, M. Livio, and S. Casertano for stimulating discussions.
I am grateful to the anonymous referee whose comments have
helped to improve the presentation of the results. }


\begin{references}

 \reference{barbier94}
Barbier-Brossat M., Petit M., Figon P. 1994, \aaps, 108, 603

 \reference{cayrel97}
Cayrel de Strobel G., Soubiran C., Friel E.D., Ralite N., \&\ Francois P.
1997, \aaps, 124, 299

 \reference{caloi99}
Caloi, V., Cardini, F., D'Antona, F., Badiali, M., Emannuele, A., \&\ 
Mazzitelli, I. 1999, \aap, 351, 925.

 \reference{carlberg85}
Carlberg, R. G., Dawson, P.C., Hsu, T., \&\ VandenBerg, D. 1985, \apj, 294, 674

 \reference{clausen99}
Clausen, J. V., Baraffe, I., Claret, A., \&\ VandenBerg, D. A. 1999,
in Theory and Tests of Convection in Stellar Structure,
ASP Conf. Ser., 173, 265 

\reference{yale96}
Demarque, P., Chaboyer, B., Guenther, D., Pinsonneault, M., Pinsonneault, L.,
and Yi, S. 1996, Yale Isochrones

\reference{duflot95}
Duflot M., Figon P., \&\ Meyssonnier N. 1995, \aaps, 114, 269

\reference{duquennoy91}
Duquennoy A. \&\ Mayor, M. 1991, \aap, 248, 485

\reference{edvard93}
Edvardsson, B., Andersen, J., Gustaffson, B., Lambert, D. L.,
   Nissen, P. E., \&\ Tomkin., J 1993, \aap, 275, 101

\reference{gomez97}
Gimenez, A., Claret, A., Ribas, I., \&\ Jordi, C. 1999,
in Theory and Tests of Convection in Stellar Structure,
ASP Conf. Ser., 173, 41 


\reference{gomez97}
Gomez, A. E., Grenier, S., Udry, S., Haywood, M., Meilon, L., Sabas, V.,
Seillier, A. \&\ Morin, D. 1997, in Proceedings of the ESA Symposium
`{\it HIPPARCOS}  Venice '97',  ESA, p. 621


\reference{hauck98}
Hauck, B. \&\ Mermilliod, M. 1998, \aaps, 129, 431

\reference{livio93}
Livio, M. 1993, in  Blue Stragglers, APS Conf. Ser., 53, 3

\reference{marsak90}
Marsakov, V. V. Shevelev, Yu. G., \&\ Suchkov., A. A. 1990, \apss, 172, 51

\reference{meusinger91}
Meusinger, H., Reimann, H.-G., \&\ Stecklum, B. 1991, \aap, 245, 74

 \reference{moon85}
 Moon, T. T.  1985, Commun. Univ. London Obs., No. 98
 
 \reference{moon85}
 Moon, T. T. \&\ Dworetsky, M.M. 1985, \mnras, 217, 305
 
\reference{ochsen83}
Ochsenbein F. 1980, Bull. Inf. CDS 19, 74

\reference{pols94}
Pols, O. R. \&\ Marinus, M. 1994, \aap, 288, 475

\reference{pols94}
Popper, D. M. 1997, \aj, 114, 1195

\reference{suchkov98}
Suchkov, A. A. 1998, \baas, 30, 1347

\reference{suchkov99}
Suchkov, A. A. 1999, \baas, 31, 1434; \apj, in preparation 

\reference{suchkov_mcm99}
Suchkov, A. A. \&\ McMaster, M. 1999, \apj, 524, L99

\reference{twarog80}
Twarog, B. A. 1980, \apj, 242, 242

\reference{wheeler79}
Wheeler, J. C. 1979, \apj, 234, 569

\end{references}
\end{document}